\documentclass[12pt]{article}
\usepackage{amssymb}
\usepackage{mathrsfs}
\usepackage{bm}
\usepackage{graphicx}
\usepackage{subfigure}
\usepackage{hyperref}
\begin{document}
\begin{titlepage}
\begin{center}
\begin{Large}
{\bf  Information processing with Page--Wootters states}
\end{Large}
\vskip1truecm
Stam Nicolis\footnote{E-Mail: Stam.Nicolis@lmpt.univ-tours.fr}

{\sl CNRS--Laboratoire de Math\'ematiques et Physique Th\'eorique (UMR 7350)\\
F\'ed\'eration de Recherche ``Denis Poisson'' (FR 2964)\\
D\'epartement de Physique\\
Universit\'e ``Fran\c{c}ois Rabelais'' de Tours\\
Parc Grandmont, Tours 37200, France}

\end{center}

\vskip1truecm

\begin{abstract}
In order to perceive that a physical system evolves in time, two requirements must be met: (a) it must be possible to define a ``clock'' and (b) it must be possible to make a copy of the state of the system, that can be reliably retrieved to make a comparison. We investigate what constraints quantum mechanics poses on these issues, in light of recent experiments with entangled photons. 

\end{abstract}
\end{titlepage}
\newpage
\section{Introduction}
A focal point of the research of John Nicolis is information processing: his work contributed in extending the scope of the concept from the purely engineering point of view to game theory, the modeling of social phenomena and biological systems. In his latest work he was particularly interested in how the quantum properties of matter and, in particular, how entangled states could be relevant to information processing by systems such as proteins. In this contribution  I would  like to present some results on a very simple model, that shed light  on the questions he liked to raise~\cite{JSN1986,JSN1991}.

The problem we want to solve is the following: if we have a ``closed'' system, i.e. isolated from its surroundings, under what circumstances can it ``tell'' time, i.e. define a clock? These questions were framed in  a particularly concrete way thirty years ago by Page and Wootters~\cite{PW} within the context of quantum gravity and, since then,   the discussion has mainly  been  pursued with these problems in mind~\cite{PWrefs}. On the other hand, the issues raised by Page and Wootters are not limited to quantum gravity. 
 In recent years technology has led to the possibility of performing experiments on entangled photons, emulating some of  the systems envisaged in ref.~\cite{PW}, cf. for instance, ref.~\cite{Morevaetal}. To understand the experiments it is useful to perform simulations. This will be the main object of the present study. 
 
 It is important  to realize that  for an isolated system without any ``internal parts'', it is impossible to define what time (or space for that matter) is and thus  what the corresponding devices, such as ``clocks'' or ``rulers'', could measure-indeed,  by definition, there aren't any. To be able to talk about time at all it's mandatory that the system either have ``internal parts'' (subsystems or more than one states) that can eventually play the role of clocks, or be in contact with another system, whose ``evolution'' can serve as reference. The movement of the Earth with respect to the Sun or the Moon is a time--honored example. 

This, however, is  not sufficient: For the evolution to be observable it is, also, necessary for a {\em copy} of the system's state(s) to be available: we need a copier and a retrieval device-an indexation mechanism-a ``memory'' and a ``processing unit''. 
 Otherwise, it is not possible to perceive any evolution at all-we ``see'' the system always in the same state, since we don't have any means of comparison. In computer terms, we need a ``buffer''. 
 
 These arguments are, of course, classical since the language we use describes classical notions. Mathematical analysis allows us to describe quantum mechanical systems. It is here that the subtleties of quantum information processing enter the picture~\cite{steane}. 
 
 Let us, therefore,  recall some features  of {\em classical} information processing~\cite{JSN1986,FeynmanLecComp} and comment on the difference with quantum information processing. 
 
 In classical information processing the fundamental unit is the {\em bit} (binary unit), $b$, that can take two possible values, 0 or 1. While modern digital computers use transistors--that work by the principles of quantam mechanics--to realize the two possible voltage differences that are mapped to the two possible 
 values of a bit, the mapping itself is classical and the bits are organized into bytes and words and manipulated according to the rules of Boolean logic using transformations, called {\em gates}. There are two gates that can act on one bit: the identity and the negation, that ``flips'' the bit, changing its value from 0 to 1 or vice versa. 
 
 Since we need a copy, $b_2$,  of the bit, $b\equiv b_1$, we are interested in the transformations, that act on two bits. We readily deduce that there are $2^2=4$ possible inputs, so there exist four possible transformations that map a state of two bits to that of two other bits.  
 
In quantum information processing~\cite{steane,FeynmanLecComp} the fundamental unit is the {\em qubit} (quantum binary unit), $|b\rangle$: This, too, can be in two possible states, $|-1\rangle$ or $|1\rangle$. It,  typically, describes the polarization state of a photon or of an electron, in more complicated situations 
it describes the intrinsic magnetic moment (spin) of an ion in a trap (and, then, can  can be in more than two possible states). 
 But quantum mechanics only allows us to compute the probability for finding the photon, for instance, in the polarization state $|1\rangle$ or the state $|-1\rangle$, corresponding to its helicity, or the spin component along any fixed axis of the electron.  The general state of a qubit is thus given by the following expression
\begin{equation}
\label{qubit}
|b\rangle = \alpha|1\rangle + \beta|-1\rangle
\end{equation}
with $\alpha$ and $\beta$ complex numbers such that $|\alpha|^2 + |\beta|^2=1$. The probability to find the qubit, $|b\rangle$ in the state $|1\rangle$ is 
$|\langle1|b\rangle|^2=|\alpha|^2$ and in the state $|-1\rangle$ is $|\langle -1|b\rangle|^2=|\beta|^2$.  We deduce that the space of all possible states a qubit can be found in is described by the equation
\begin{equation}
\label{bloch-sphere}
|\alpha|^2+|\beta|^2=1
\end{equation}
which is the equation of  the 3--sphere $S^3$ of unit radius. However only the relative phase of the two complex numbers, $\alpha$ and $\beta$, is physically relevant: if we multiply both by the same complex number, $\exp(\mathrm{i}\theta)$, of unit modulus, we get the same point on the sphere. Therefore only two degrees of freedom remain: the space is $S^3/U(1)=S^2$, a 2--sphere, called the {\em Bloch sphere}. The two poles of this sphere correspond to the possible states of the (classical) bit. 

The time evolution of the single qubit should, also, respect this relation, expressing thereby the fact that the qubit is, indeed, a closed system. 

When we wish to couple qubits together,  the quantum analogs of gates will be transformations on the space of qubits, in the same way that classical gates are transformations on the space of bits. Whereas we can only flip the state of a single bit, through the negation, a qubit has a much richer set of transformations, namely those that realize a motion of a point on the unit 2--sphere and leave the origin fixed and the radius equal to unity, since it represents the probability of finding the qubit in any possible state. 

When we consider the analogs of the gates that realize transformations of two qubits we therefore come to the conclusion that they, too, must obey such a constraint.  We may write the corresponding transformation as follows:
\begin{equation}
\label{2qubits}
|\psi\rangle=|b_1'\rangle|b_2'\rangle = c_1|-1\rangle|-1\rangle + c_2|-1\rangle|1\rangle + c_3|1\rangle|-1\rangle + c_4|1\rangle|1\rangle
\end{equation}
where the $c_i$, $i=1,2,3,4$ are complex numbers, that satisfy $|c_1|^2+|c_2|^2+|c_3|^2+|c_4|^2=1$. Once more we remark that we may multiply all the $c_i$ by the same complex number, of unit modulus, $\exp(\mathrm{i}\theta)$, without affecting this relation, that expresses, once more, conservation of probability. 
This describes a 7--sphere in an eight--dimensional space, but we must identify the points that are related by a global phase, so the space of configurations of all possible pairs of qubits is $S^7/U(1)$, a much more complicated space to describe than the square, the space of configurations of all possible pairs of bits. 

The phase of the complex numbers $c_i$ is the relative phase of the two qubits.

Now that we have described the space of 2--qubit configurations, we can focus on the problem of characterizing the transformations of this space. These can then be identified as one--step evolution operators.  We wish to classify those evolution operators that have the property that, when they act on a 2--qubit state, the relative phases do not all change by the same amount. This is how the system will be able to register the passage of time. 

This is the subject of the next section, where we recall  the idea underlying the proposal by Page and Wootters and that of the experiment of ref.~\cite{Morevaetal} and  set them within this classification scheme.   
 
 In section~\ref{Simulations} we describe some simulations that highlight our approach.  We close with our conclusions and a discussion of directions of further inquiry.  
 
 \section{A generalized Page--Wootters framework for two photons}\label{PWframework}
  The analysis of ref.~\cite{PW} considered a closed system of $N$ spin$-j$ particles.  For simplicity we will rather deal here with systems of two photons, considered in  the experiment of ref.~\cite{Morevaetal}  The states of the system are the product kets, $| m_1\rangle|m_2\rangle$, with $m_I=\pm 1$, the helicities of the photons. The system has four states in total so we can talk, meaningfully, about transition amplitudes between them. 
  
Since the system is closed  the evolution operator, whose expectation values are the transition amplitudes, must be a unitary  operator~\cite{schwinger}.  A one-step evolution operator of this kind is given by the following expression
\begin{equation}
\label{evolop}
U_{k,l}= P_{k,l}
\end{equation}  
where $P_{k,l}=\delta_{k,l+1}$ is the one-step shift operator in the space of the four states. This simply permutes the states and   was used in ref.~\cite{Morevaetal}. It has, of course, the property that $U^4=I_{4\times 4}$. 

But this operator is  not the only one allowed. The most general unitary operator that acts on the space of these four states  can be constructed by introducing the operator $Q_{k,l}=\exp(2\pi\mathrm{i}k/4)\delta_{k,l}\equiv\omega_4^k\delta_{k,l}$, where $\omega_4\equiv\exp(2\pi\mathrm{i}/4)$ is the fourth root of unity. Then the operator~\cite{balian_itzykson} 
\begin{equation}
\label{Uevolop}
U(\phi)_{k,l}=c(\phi)\sum_{r,s=0}^3 e^{2\pi\mathrm{i}\phi_{r,s}/4}\omega_4^{rs/2}\left[P^rQ^s\right]_{k,l}
\end{equation}
with $\phi_{r,s}$ real is  a unitary operator. 
Since the operators $P$ and $Q$ do not commute, the action of $U(\phi)$ on a state vector is not, simply, a permutation. Nonetheless, the system remains closed and an ``outside observer'', being unable to measure the global phase,  observes a static system. 

The operator of eq.~(\ref{Uevolop}) also   satisfies $U(\phi)^4=I_{4\times 4}$. Another, well--known, member of this family is the Discrete Fourier Transform (DFT) for two qubits,
\begin{equation}
\label{DFT}
F_{k,l}=\frac{\omega_4^{kl}}{2}
\end{equation}
The property that will interest us here is that it doesn't commute with $P$, indeed its columns are the eigenvectors of $P$: $F P  = QF$. 

The reason this property is desirable is that we want  to use one photon as a clock for the other. To achieve this we only have the relative phase of the two photons at our disposal and it is easy to prove that, under $U_{k,l}=P_{k,l}$, the relative phase of the two photons does not change, upon each application of the evolution operator. 

We must thus introduce a mechanism to make the relative phases change. The experiments done in ref.~\cite{Morevaetal} use quartz slabs to induce a phase difference. The length of the slabs is the proxy for the number of times the evolution operator acts. This defines the ``clock'' within the two--photon system. (We remark that they thereby ``couple'' the, original, two-photon system to an external device.)

One way to model this theoretically, while remaining within the two--photon system,  is by using the operator $U(\phi)$ under whose action  the relative phases do not stay fixed, since $P$ and $Q$ do not commute. Therefore, with the operator $U(\phi)$ as evolution operator we {\em can} define a clock and the system will be able to ``tell'' time, by comparing the relative phases between the states. 

Let us show how the system can tell time, when we use $U(\phi)=F$, the Discrete Fourier Transform.

 \section{Simulations}\label{Simulations}
 We start with the state 
 \begin{equation}
 \label{initialstate}
 |\psi\rangle = c_1|-1\rangle|-1\rangle + c_2|-1\rangle|1\rangle + c_3|1\rangle|-1\rangle + c_4|1\rangle|1\rangle
 \end{equation}
 normalized to unity, $|\langle \psi|\psi\rangle|^2=1$. 
 Ref.~\cite{Morevaetal}, for instance,  take $c_2=0=c_3$ and $c_1=\cos\omega, c_4=\sin\omega$, with $\omega$ some angle. 
 We want to compute $U(\phi)^n|\psi\rangle\equiv|\psi_n\rangle$ and show that $|\langle\psi|\psi_n\rangle|^2$ does depend on the index $n$, therefore there does exist a relative phase between the two photons, that is sensitive to the number of times the operator $U(\phi)$ has acted on the system as a whole. 
 Since $U(\phi)^4=I_{4\times 4}$, the physically interesting quantity is $n\,\mathrm{mod}\,4=0,1,2,3$. 
   
 To provide a flavor, we shall choose as one-step evolution operator, $U(\phi)\equiv F$, the Discrete Fourier Transform of order 4. We  may compute $U^n|\psi\rangle$, by expanding $|\psi\rangle$ on the eigenvectors of $F$ for instance~\cite{mehta}.  For the case at hand we note that $F^n=F^{4m+k}=F^k$, with 
 $k=0,1,2,3$. So we just need to compute the action of $F$ and of $\left[F^2\right]_{k,l}=\delta_{k,-l}$, which is the parity operator, on the initial vector. To do that we use the correspondence 
\begin{equation}
\label{correspondence}
\begin{array}{lcl}
|-1\rangle|-1\rangle &\leftrightarrow &(1,0,0,0)^\mathrm{T}\\
|-1\rangle|1\rangle   &\leftrightarrow &(0,1,0,0)^\mathrm{T}\\
|1\rangle|-1\rangle   &\leftrightarrow &(0,0,1,0)^\mathrm{T}\\
|1\rangle|1\rangle    &\leftrightarrow &(0,0,0,1)^\mathrm{T}\\
\end{array} 
\end{equation} 
to define how the one-step evolution operator acts on the states. 

\begin{figure}[thp]
 \begin{center}
 \includegraphics[scale=0.9]{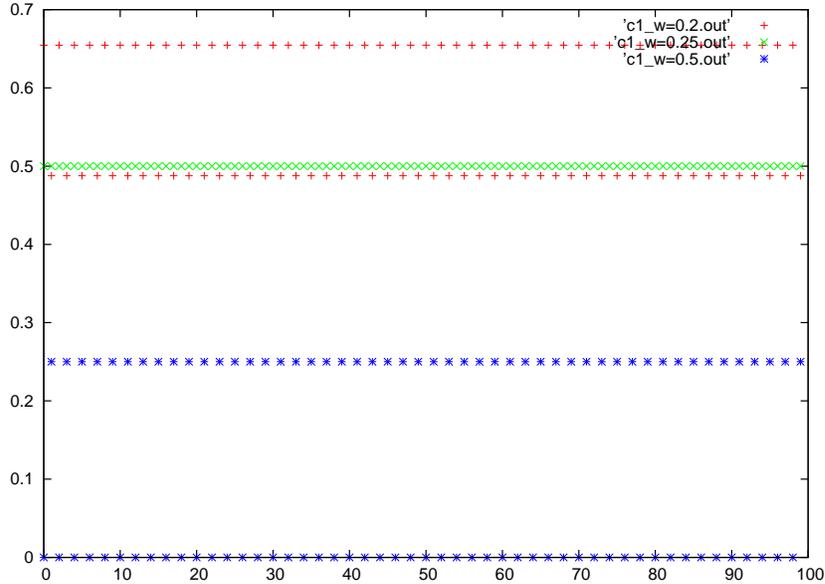}
 \end{center}
 \caption[]{Plot of $|c_1(n)|^2$ vs. the time step $n$ for $\omega/\pi=0.2,0.25,0.5$. For $\omega/\pi=0.25$ there is just one line of points, for the others two.}
 \label{FourierU}
\end{figure}

 We find the following result, 
 \begin{equation}
 \label{Unpsi}
|\psi_n\rangle = F^n|\psi\rangle=F^k|\psi\rangle\equiv c_1(k)|-1\rangle|-1\rangle + c_2(k)|-1\rangle|1\rangle + c_3(k)|1\rangle|-1\rangle + c_4(k)|1\rangle|1\rangle
 \end{equation} 
 where ($c_1(0)=\cos\omega\equiv c, c_2(0)=0=c_3(0),c_4(0)=\sin\omega\equiv s$)
 \begin{equation}
 \label{ckcoeffs}
 \begin{array}{lll}
c_1(1)= (c+s)/2 & c_1(2)= c& c_1(3)=(c+s)/2\\
c_2(1)= e^{-\mathrm{i}\omega}/2& c_2(2)= s& c_2(3)=e^{\mathrm{i}\omega}/2\\
c_3(1)=(c-s)/2& c_3(2)= 0& c_3(3)=(c-s)/2 \\
c_4(1)= e^{\mathrm{i}\omega}/2& c_4(2)=0 & c_4(3)=e^{-\mathrm{i}\omega}/2\\
 \end{array}
 \end{equation}
 Unitarity of the evolution operator implies that $|c_1(n)|^2+|c_2(n)|^2+|c_3(n)|^2+|c_4(n)|^2=1$, implying, also, that the system, as a whole, is perceived as ``static'' by an ``external'' observer. Therefore the interesting quantities are the $|c_i(n)|^2$. If these are independent of $n$, then we cannot use the corresponding states to tell time. For example, if $\omega=\pi/4$, then $c=s=1/\sqrt{2}$ and $|c_1(n)|^2=1/2$. The relative phase  of the two photons does not change with time in this state. To tell time, therefore, if one photon's helicity is +1, the other's helicity must be $-1$, if $\omega=\pi/4$. In fig.~\ref{FourierU} 
 we display $|c_1(n)|^2= |\langle -1|\langle -1|\psi_n\rangle|^2$ as a function of the time step $n$, for different values of $\omega$. For $\omega\neq\pi/4$ we find two lines, for $\omega=\pi/4$ just one. The two values, for $\omega\neq\pi/4$, indicate that the system can tell time and one, also, finds that the two values are visited periodically. 

Let us draw attention now to the following point: If we start with an initial qubit configuration, $|\psi_0\rangle$ and wish to compute $|\psi_1\rangle = F|\psi_0\rangle$, when we use the Fast Fourier Transform we replace $|\psi_0\rangle$ by $|\psi_1\rangle$. This precludes, therefore, the possibility of computing $|\langle\psi_0|\psi_1\rangle|^2$. In order to perform this calculation we must make a copy of $|\psi_0\rangle$. 

On a classical computer this isn't an issue of principle, but it does require allocating the corresponding memory: Each qubit pair is represented classically by the array of complex numbers $c_i$, on which acts the evolution operator, a $4\times 4$ matrix with complex entries. So we just need to keep a copy of the initial vector and then exchange the current vector for the previous one, for the evolution. 

On a quantum computer, however, making a copy of $|\psi_0\rangle$ means finding a unitary operator, $C$ that  performs the following operation~\cite{copying}
\begin{equation}
\label{CopyU}
C|\psi_0\rangle = |\psi_0\rangle|\psi_0\rangle|c\rangle
\end{equation}
 where $|c\rangle$ is the state of the copier, which must hold, of course, at least the qubit configuration being copied.  The ``no-cloning theorem'' ~\cite{copying} states that such an operator does not exist. What can be constructed~\cite{copying} is a copying operator that performs a copy with some finite precision, which implies an accumulation of error in the copying with time, that certainly deserves to be studied in detail. 
 
  We shall not enter into the details of the copying procedure, but refer to the paper~\cite{copying} for the technical details themselves. A detailed study will be presented in future work.  
 
 This is the result that does not take the entanglement due to copying into account. It does illustrate that two photons can ``tell'' time, if their dynamics is rich enough, even though they are a closed system. 

 The ``problem of time'', in fact, is the ``problem of finding a copier''. And this problem, at the quantum level, while non-trivial, since the copier is not perfect, has a well-defined solution. What we haven't had the opportunity to discuss is how to ``erase'' a copy, in order to {\em replenish} the buffer. This raises a host of further issues that need to be brought together~\cite{dipab}. While the problem of ``erasure'' is at the heart of quantum computing~\cite{querase}, the many, seemingly disparate, issues remain to be brought into a broader synthesis, an element that was, also at the heart of refs.~\cite{JSN1986,JSN1991}.

 \section{Conclusions}\label{Conclusions}
 In conclusion we have presented an example of a closed system that can ``tell'' time operationally,  in a fully quantum mechanical way. The new feature that we needed,  to do the job, was the buffer memory. Quantum mechanics imposes severe constraints on its use, due to the non-cloning theorem. Nonetheless, transition amplitudes between the finite number of states of the system can be defined and we obtain, in this way, a simulation of the experiment presented in ref.~\cite{Morevaetal} that eliminates the need for many of the assumptions introduced there. In particular, we remark that unitarity of evolution of the full system ensures that the ``center of mass energy'' remains an unobservable, global phase--there is no need or way to set it to zero, once we have defined the cocycle properly. We realize that we may use a much larger class of evolution operators than hitherto considered and hope to expand on this in further work.
 
 As a bonus we acquire some insights into the information processing capabilities of closed quantum systems that may, also,  serve as a starting point for investigating open quantum systems. Such insights might, in particular, prove useful for understanding the quantum information processing by black hole microstates~\cite{AFNModDisc}.

\end{document}